\documentclass{PoS}

\title{Top-Quark Asymmetry -- A New Physics Overview}

\ShortTitle{Top-Quark Asymmetry -- A New Physics Overview}

\author{Susanne Westhoff\\
Institut f\"ur Physik (THEP) and Helmholtz-Institut Mainz\\
Johannes Gutenberg-Universit\"at \\
D-55099 Mainz, Germany \\
        E-mail: \email{westhoff@uni-mainz.de}}

\abstract{Several recent measurements at the Tevatron experiments point towards an anomalously large forward-backward asymmetry in top-antitop production. This article summarizes the main classes of physics beyond the standard model that can give rise to a large asymmetry. Complementary measurements at the Large Hadron Collider will allow to distinguish between different models and thereby contribute to clarifying the presently puzzling picture.}

\FullConference{International Europhysics Conference on High Energy Physics -- HEP 2011\\
		July 21-27, 2011\\
		Grenoble, Rh\^{o}ne-Alpes, France\hfill \rm{MZ-TH/11-24}}

\begin{document}
\section{New physics in top-quark pair production}
\begin{figure}[t]
\begin{center}
\vspace*{-0.5cm}
\includegraphics[height=6.7cm]{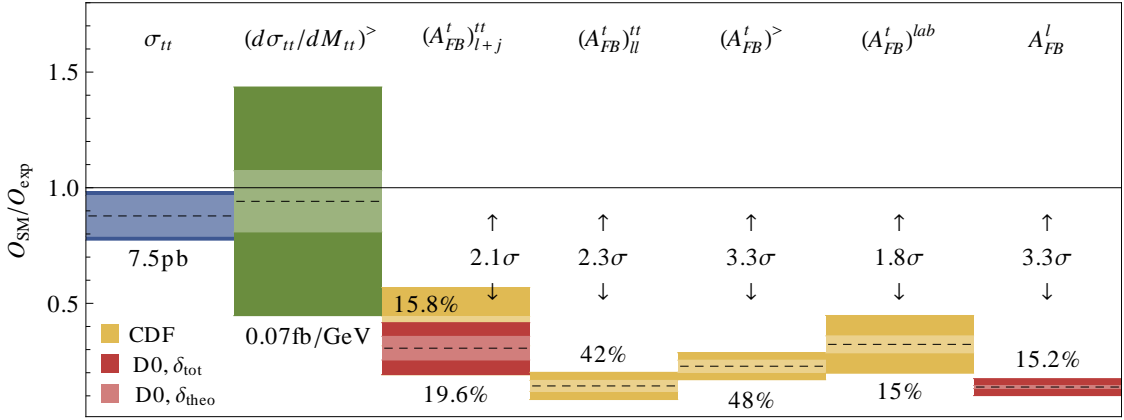}
\parbox{15.5cm}{\caption{\label{fig:SMexp}Top-antitop production at the Tevatron. The ratio $O_{\rm{SM}}/O_{\rm{exp}}$ is displayed for the total cross section $\sigma_{t\bar t}$ and
its invariant mass distribution $(d\sigma/dM_{t\bar t})^>$ for $M_{t\bar t} \in [0.8, 1.4]$ TeV. The inclusive asymmetry in the parton frame is shown for the lepton + jets channel, $(A_{\rm{FB}})^{t\bar t}_{l+j}$, besides its bin $(A_{\rm{FB}}^t)^>$ for high invariant mass $M_{t\bar t} > 0.45$ TeV, as well as for the dilepton channel, $(A_{\rm{FB}}^t)^{t\bar t}_{ll}$. The asymmetry in the laboratory frame is denoted by $(A_{\rm{FB}}^t)^{\rm{lab}}$, and $A_{\rm{FB}}^l$ is the charged lepton asymmetry. Numbers correspond to the central measured values \cite{Aaltonen:2011kc}.}}
\end{center}
\end{figure}

The anomaly in top-quark pair production persists: both the CDF and D\O\ experiments at the Tevatron have measured the forward-backward asymmetry $A_{\rm{FB}}^t$ of the top quark \cite{Aaltonen:2011kc} and consistently found higher results compared to the standard model (SM) prediction. The present situation is illustrated in Figure~\ref{fig:SMexp}. While the top-antitop production cross section $\sigma_{t\bar t}$ and its invariant mass spectrum $d\sigma_{t\bar t}/dM_{t\bar t}$ agree well with the SM predictions, the asymmetric observables exhibit deviations of up to $3\sigma$ significance. In particular, CDF observes a significant forward-backward excess at large invariant mass, which is less pronounced at D\O, but agrees with the CDF result at the level of reconstructed data within one standard deviation. In the framework of Quantum Chromodynamics (QCD), the forward-backward asymmetry is a suppressed quantity, arising only at next-to-leading order in powers of the strong coupling constant \cite{Kuhn:1998kw}. The small QCD value $(A_{\rm{FB}}^t)^{\rm{lab}}_{\rm{SM}}=(4.8\pm 0.5)\%$ \cite{Ahrens:2011uf} is expected to be robust with respect to higher-order corrections \cite{Ahrens:2011uf,Melnikov:2010iu} and increased by electroweak contributions by at most $20\%$ \cite{Hollik:2011ps}.

\begin{figure}[t]
\begin{center}
\includegraphics[height=2.5cm]{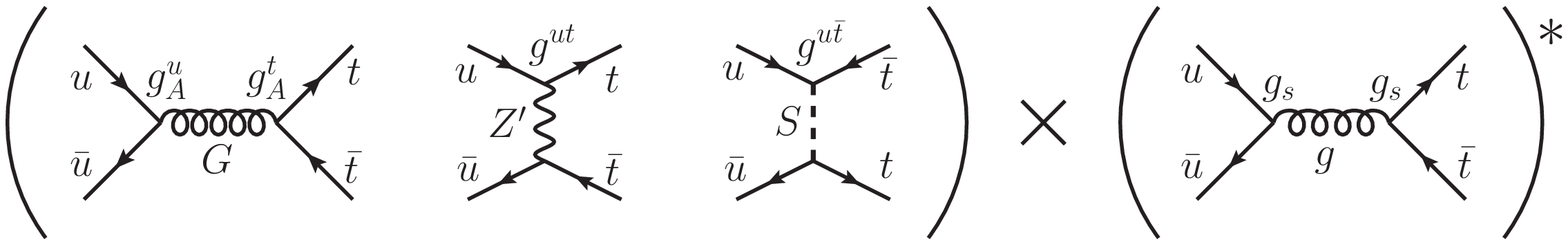}
\parbox{15.5cm}{\caption{\label{fig:feynman}}Interference of new particles in the $s$, $t$, and $u$ channels of the process $u\bar u\rightarrow t\bar t$ with the SM gluon.}
\end{center}
\end{figure}

Due to its smallness in QCD, the asymmetry is not only a powerful test of the QCD vector current, but also a sensitive probe of physics beyond the SM. In a theory with CP-conserving couplings, the forward-backward asymmetry equals a charge asymmetry, defined by
\begin{equation}
 A_{\rm{FB}}^t \equiv A_c^t = \frac{\sigma_a}{\sigma_s}\,,\qquad \sigma_{a(s)} = \int_0^1 \cos\theta\left[\frac{d\sigma(p\bar p\rightarrow t\bar t X)}{d\cos\theta} -(+) \frac{d\sigma(p\bar p\rightarrow \bar t t X)}{d\cos\theta}\right]\,,
\end{equation}
where $\theta$ is the scattering angle of the top quark with respect to the direction of the incident proton $p$. (New) contributions to $A_c^t$ thus stem from a partonic $q\bar q\rightarrow t\bar t$ amplitude that is odd under the interchange of the top with the antitop quark in the final state. The process $gg\rightarrow t\bar t$ is symmetric in its initial state and therefore does not yield an asymmetry. Since a large positive effect in $A_{\rm{FB}}^t$ is required to explain the discrepancy, it is most likely due to the interference of the SM gluon with a new particle exchanged at tree level. This statement is also supported by the pattern of $t\bar t$ data at high invariant mass $M_{t\bar t}$  \cite{Grinstein:2011yv}, where the anomaly in the asymmetry is most significant. In a first approach to classifying those interference effects, one can work in the framework of an effective theory \cite{Delaunay:2011gv,AguilarSaavedra:2011vw}.\footnote{As the scale of new physics in $A_{\rm{FB}}^t$ is typically in the range of the momentum transfer in $q\bar q\rightarrow t\bar t$, width and resonance effects, which are not covered in an effective theory, can be important.} Among the dimension-six four-quark operators mediating the dominant $u\bar u\rightarrow t\bar t$ transition, the relevant operators that interfere with the QCD amplitude are
\begin{equation}
 \begin{array}{llll}
\mathcal{O}_V^8 &= (\bar u\,\gamma_{\mu}T^a\,u)(\bar t\,\gamma^{\mu}T^a\,t)\,, & \quad \mathcal{O}_A^8 &= (\bar u\,\gamma_{\mu}\gamma_5 T^a\,u)(\bar t\,\gamma^{\mu}\gamma_5 T^a\,t)\,,\\[0.1cm]
\mathcal{O}_S^3 &= (\bar t\,T_3\,u)(\bar u\,T^3\,t)\,, & \quad \mathcal{O}_P^3 &= (\bar t\,\gamma_5 T_3\,u)(\bar u\,\gamma_5 T^3\,t)\,.\label{eq:operators}
\end{array}
\end{equation}
Here $T^a$ are the generators of the QCD color octet, and $T_3=\epsilon_{abc}$ is the antisymmetric tensor of a color triplet. We will distinguish the tree-level contributions of these operators to the process $u\bar u\rightarrow t\bar t$ in terms of Mandelstam variables. The operators $\mathcal{O}_V^8$ and $\mathcal{O}_A^8$ mediate vector and axial-vector currents in the $s$ channel, while flavor-changing chiral currents in the $t$ channel arise after a Fierz transformation. The operators $\mathcal{O}_S^3$ and $\mathcal{O}_P^3$ induce flavor-changing scalar and pseudo-scalar currents in the $u$ channel. The corresponding Feynman diagrams are shown in Figure~\ref{fig:feynman}. As we will discuss in detail, there are three main mediators of large contributions to $A_{\rm{FB}}^t$: a color-octet vector boson with axial-vector couplings $g_A^q$ to quarks in the $s$ channel, a color-singlet vector in the $t$ channel or a color-triplet scalar in the $u$ channel, both with flavor-changing couplings $g^{ut}$, $g^{u\bar t}$ to quarks.

In the following, we will review the three classes of new physics (NP) in top-antitop production. Our main focus will be on models that reconcile the theory prediction of the asymmetry with its measurement, while preserving the good description of the $t\bar t$ cross section and its $M_{t\bar t}$ distribution in terms of QCD. Further strong constraints from flavor physics and collider observables have to be respected when constructing a viable model. For each class, we give examples of concrete realizations of NP that yield a consistent picture of top-quark pair production.

\section{s channel: color-octet vectors (axigluons)}
Color-octet vector bosons with axial-vector couplings $g_A^q$ to quarks, dubbed axigluons, generate a charge asymmetry at tree level. The interference of an axigluon with the QCD gluon (INT) and the interference of an axigluon with itself (NP) yield the following contributions to $t\bar t$ production,
\begin{equation}
\sigma_a^{\rm{INT}}\sim g_s^2\frac{g_A^q\,g_A^t}{M_{t\bar t}^2-M_G^2}\,,\qquad \sigma_s^{\rm{NP}} \sim (g_A^q)^2(g_A^t)^2\frac{M_{t\bar t}^2}{(M_{t\bar t}^2-M_G^2)^2}\,,
\end{equation}
where $M_G$ is the mass of the axigluon and $g_s$ denotes the strong coupling constant of QCD. The sign of the interference term $\sigma_a^{\rm{INT}}$ is determined by the magnitude of the axigluon mass with respect to the momentum transfer in the process $q\bar q\rightarrow t\bar t$ and by the signs of the quark couplings $g_A^q$ and $g_A^t$. In order to generate a positive asymmetry, a light axigluon of $M_G\approx M_{t\bar t}$ may have flavor-universal couplings to quarks, whereas a heavy axigluon with $M_G\gg M_{t\bar t}$ must exhibit axial-vector couplings of different sign to light and top quarks. A strong constraint on the magnitude of $g_A^q g_A^t/M_G^2$ arises from the $t\bar t$ production cross section. Even though the axigluon does not interfere with the charge-symmetric QCD amplitude, the NP term $\sigma_s^{\rm{NP}}$ causes significant distortions in the invariant mass spectrum, in particular close to the resonance region.

Indirect constraints on axigluon effects in $t\bar t$ production arise from flavor physics and electroweak precision observables. In the case of flavor non-universal couplings, the misalignment of the axigluon couplings with the Yukawa couplings induces flavor-changing neutral currents (FCNC) at tree level. In the absence of a flavor theory, FCNC effects can be confined to the up-type quark sector \cite{Bai:2011ed,Haisch:2011up}. $D-\overline{D}$ meson mixing then constrains the mass of an axigluon with quark couplings of QCD strength to be beyond the electroweak scale, $M_G \gtrsim 200\,\rm{GeV}$ \cite{Haisch:2011up}. Constraints from electroweak precision observables are due to one-loop axigluon corrections to the $Zq\bar q$ vertex. The precise measurements of the $Z$ boson width, its hadronic cross section and its coupling to bottom quarks at LEP yield a lower bound on the axigluon mass of $M_G\gtrsim 500\,\rm{GeV}$ \cite{Haisch:2011up}. Axigluon contributions to the electroweak gauge boson propagators, encoded in the oblique parameters $S$ and $T$, occur first at the two-loop level. The mass constraints are thus rather mild, but reach the level of several $100\,$GeV for strong axial-vector couplings to top quarks \cite{Haisch:2011up}.

A serious limiting factor of axigluon effects in top-quark pair production is their simultaneous contribution to dijet production at the LHC. Both ATLAS and CMS have carried out searches for resonances in the dijet spectrum from proton-proton collisions at a center-of-mass energy of $\sqrt{s}=7\,\rm{TeV}$, as well as for anomalies in the dijet angular distribution. The agreement with the QCD predictions strongly constrains the mass of an axigluon with QCD-like couplings to $M_G\gtrsim 2\,\rm{TeV}$ \cite{Bai:2011ed,Haisch:2011up}. In general, dijet bounds can be evaded by decreasing the couplings to light quarks, while keeping the relevant product for the asymmetry, $g_A^q g_A^t$, fixed.\footnote{When increasing the top-quark couplings, however, constraints from the $T$ parameter should be kept in mind \cite{Haisch:2011up}.}

A broad class of concrete axigluon models relies on the concept of chiral color \cite{Frampton:1987dn}, an extension of the QCD gauge group $SU(3)_C$ to the chiral product $SU(3)_L\times SU(3)_R$. The massive axigluon arises once the product group is spontaneously broken to the diagonal subgroup $SU(3)_C$. In this framework, the origin of the axigluon restricts the couplings to light quarks and top quarks to be of equal strength, $|g_A^q|=|g_A^t|$. A heavy axigluon with $M_G=2\,\rm{TeV}$, $g_A^q = -g_A^t = 1$ and a width $\Gamma_G/M_G=10\%$ evades flavor and collider constraints and yields a top-quark asymmetry at high invariant masses of up to $(A_{\rm{FB}}^t)_{\rm{SM+NP}}^> = 20\%$ \cite{Haisch:2011up,Ferrario:2009bz}. The effect is limited by the bounds from dijet production. Light axigluons should have been observed as resonances in $t\bar t$ production, as well as dijet production. A broad width can hide an axigluon resonance, but requires new states in the axigluon decay \cite{Barcelo:2011vk}. An attractive solution to the top asymmetry anomaly is a light axigluon with $M_G \approx 400\,\rm{GeV}$, universal couplings $g_A^q = g_A^t = 1/3$, and a broad width \cite{Tavares:2011zg}. It can induce a large high-$M_{t\bar t}$ asymmetry of $(A_{\rm{FB}}^t)_{\rm{NP}}^>\approx 30\%$.

In Randall-Sundrum models, axigluons are naturally incorporated as Kaluza-Klein excitations of the gluon. In models with flavor anarchy, the mass hierarchy and mixings of quarks can be explained by their localization in the warped extra dimension: heavy quarks typically reside in the infra-red (IR), light quarks in the ultra-violet (UV) region. The IR-localized Kaluza-Klein gluons exhibit axial-vector couplings to quarks, depending on the overlap of their wave function profiles. Axial-vector couplings to the heavy top quarks are large, but strongly suppressed for the UV-localized light quarks. Without relaxing the principle of flavor anarchy, it is therefore not possible to generate a sizeable forward-backward asymmetry from Kaluza-Klein gluons \cite{Bauer:2010iq}. Moving the light quarks towards the IR to increase their axial-vector couplings, however, requires an additional flavor protection mechanism to inhibit too large effects in flavor observables \cite{Delaunay:2011vv}.

\section{t channel: color-singlet vectors ($Z^{\prime}$)}
A color-singlet vector boson, referred to as $Z'$ in analogy with the SM $Z$ boson, contributes to top-antitop production in the following way,
\begin{equation}
 \sigma_a^{\rm{INT}} \sim a\cdot g_s^2\frac{(g_L^{ut})^2+(g_R^{ut})^2}{t-M_{Z'}^2}\,,\quad \sigma_s^{\rm{INT}} \sim b\cdot g_s^2\frac{(g_L^{ut})^2+(g_R^{ut})^2}{t-M_{Z'}^2}\,,\quad \sigma_s^{\rm{NP}} \sim \frac{(g_{L,R}^{u\bar u})^2(g_{L,R}^{t\bar t})^2\,M_{t\bar t}^2}{(M_{t\bar t}^2-M_{Z'}^2)^2}\,.
\end{equation}
The exchange of a $Z'$ boson with flavor-changing couplings $g_{L,R}^{ut}$ in the $t$ channel yields a Rutherford enhancement of the top asymmetry at high invariant mass \cite{Jung:2009jz}, as observed by CDF. Since the interference with the SM gluon, $\sigma_a^{\rm{INT}}$, is negative, large couplings $g_{L,R}^{ut}$ are required to achieve a positive asymmetry from the NP term $\sigma_a^{\rm{NP}}$. Contrary to new particles in the $s$ channel, a flavor-violating boson in the $t$ channel does not appear as a resonance in the spectrum of the $t\bar t$ cross section. However, the Rutherford enhancement of $\sigma_{t\bar t}$ via the interference term $\sigma_s^{\rm{INT}}$ constrains the $Z'$ mass $M_{Z'}$ to be below a few $100\,\rm{GeV}$. The contribution $\sigma_s^{\rm{NP}}$ to the cross section prohibits large flavor-diagonal couplings $g_{L,R}^{u\bar u}$, $g_{L,R}^{t\bar t}$, which restricts the flavor structure of viable $Z'$ models.

Indirect constraints on flavor-violating $Z'$ bosons arise from meson mixing. The $SU(2)_L$ gauge symmetry of the SM relates the $u_L - t_L - Z'$ coupling to a $d_L - b_L - Z'$ coupling, which induces tree-level effects in $B_d-\overline{B}_d$ mixing. The resulting strong bound $g_L^{ut} < 3.5\times 10^{-4}(M_{Z'}/100\,\rm{GeV})$ \cite{Cao:2010zb} suggests large right-chiral $Z'$ couplings. The coupling $u-c-Z'$, connected to $u-t-Z'$ by the $SU(3)_U$ flavor symmetry, is constrained by effects in $D-\overline{D}$ mixing \cite{Jung:2009jz}. A large top-quark asymmetry could thus be due to a $Z'$ boson with a large $g_R^{ut}$ coupling, breaking both the chiral and $SU(3)_U$ flavor symmetries. In models of topcolor-assisted technicolor, top-quark condensates generate the desired large $g_R^{ut}$ coupling and provide a consistent explanation of the asymmetry \cite{Cui:2011xy}.

Severe direct constraints on $u-t-Z'$ couplings stem from same-sign top production. The recent search for $pp\rightarrow tt$ and $pp\rightarrow ttj$ processes by CMS excludes the region of the $Z'$ parameter space that is compatible with $A_{\rm{FB}}^t$ to $95\%$ confidence level \cite{Chatrchyan:2011dk}. The abelian $Z'$ model is thus ruled out by same-sign top production at the LHC. A possible way out is to protect the model from same-sign tops by a non-abelian $SU(2)^{ut}_R$ flavor symmetry. The forward-backward asymmetry in this setup is induced by the $t$-channel exchange of (electrically neutral) $W'$ gauge bosons, which carry ``flavor isospin''. This flavor quantum number forbids same-sign top production and thereby allows for a large asymmetry $(A_{\rm{FB}}^t)^>\approx 30\%$ for $M_{W'} = 200\,\rm{GeV}$ \cite{Jung:2011zv}.

\section{u channel: color-triplet scalars ($S$)}
Besides the vector bosons discussed in the previous sections, there exist two scalar fields that can yield a sizeable top-quark asymmetry: a color triplet (cf. Eq.~(\ref{eq:operators})) and a color sextet. Their quantum number assignments under the SM gauge group are
\begin{equation}
(SU(3)_C,SU(2)_L)_{U(1)_Y} :\quad S_{\overline{3}} = (\overline{3},1)_{4/3}\,,\quad S_6 = (6,1)_{4/3}\,.
\end{equation}
The color representation dictates the flavor symmetry of the scalars. The color sextet $S_6$ is also a (symmetric) sextet representation of the flavor group $SU(3)_U$, which implies flavor-diagonal couplings to quarks. A significant contribution of $S_6$ to the forward-backward asymmetry is therefore ruled out by $s$-channel dijet production via $uu \rightarrow S_6 \rightarrow uu$. The color triplet $S_{\overline{3}}$, however, has antisymmetric flavor-triplet couplings to quarks, which protects it from the dominant flavor-diagonal contributions to dijet production. The $SU(3)_U$ symmetry further prevents the scalars from inducing undesirable flavor-violating FCNC \cite{Ligeti:2011vt}.

The exchange of a color triplet $S_{\overline{3}}$ in $t\bar t$ production proceeds via the $u$ channel. The effect on the top-quark asymmetry is not Rutherford-enhanced and therefore generically smaller than for $t$-channel NP \cite{Shu:2009xf}. A clear advantage of the $u$-channel exchange is the absence of same-sign tops due to the conservation of electric charge. Light color triplets $S_{\overline{3}}$ may originate from an extended Higgs sector in models with grand unification \cite{Dorsner:2009mq}.

\section{Model distinction at the LHC}
The origin of the Tevatron forward-backward asymmetry can be clarified by exploring correlated observables at the LHC. While a forward-backward asymmetry is not observable in proton-proton collisions due to the symmetric initial state, the origin of the quark and antiquark inside the protons can be determined by their different momentum distributions. One can probe charge-asymmetric contributions to $t\bar t$ production via an asymmetry based on pseudo rapidities $\eta = -\ln(\tan\theta/2)$ \cite{Antunano:2007da},
\begin{equation}
A_{\eta}^t = \frac{N(|\eta_t| > |\eta_{\bar t}|)-N(|\eta_t| < |\eta_{\bar t}|)}{N(|\eta_t| > |\eta_{\bar t}|)+N(|\eta_t| < |\eta_{\bar t}|)}\,,\qquad (A_{\eta}^t)_{\rm{SM}} = (1.30\pm 0.11)\%\,.
\end{equation}
The CMS and ATLAS collaborations have measured $A_{\eta}^t$ consistent with zero \cite{cms:2011conf}. The results are still limited by large statistical uncertainties; systematics are expected to shrink to $(1-2)\%$. The candidates for a large $A_{\rm{FB}}^t$ typically enhance $(A_{\eta}^t)_{\rm{SM}}$ by a factor of about two. It might be difficult to distinguish new physics from the SM using $A_{\eta}^t$ without setting further cuts on $\eta_t$ and $M_{t\bar t}$ \cite{Hewett:2011wz,Arguin:2011xm}.

A promising road to discriminate between different models is to explore the $M_{t\bar t}$ spectrum of top-pair production at the LHC \cite{Hewett:2011wz,Gresham:2011fx}: an axigluon of $M_G=\mathcal{O}(1\,\rm{TeV})$ should show up as a resonance, while a flavor-violating $Z'$ boson enhances the tail of the spectrum. The chiral couplings of the top quark can be disentangled from its polarization \cite{Krohn:2011tw}. An improved measurement of the $M_{t\bar t}$ spectrum of the forward-backward asymmetry at the Tevatron is possible and will help to distinguish between models that yield the same inclusive asymmetry \cite{AguilarSaavedra:2011ci}.

\end{document}